\documentstyle[12pt,epsf,epsfig]{article}
\def\ga{\mathrel{\raise.3ex\hbox{$>$\kern-.75em\lower1ex\hbox{$\sim$}}}}
\def\la{\mathrel{\raise.3ex\hbox{$<$\kern-.75em\lower1ex\hbox{$\sim$}}}}
\def\gyr{{\rm \, G\kern-0.125em yr}}
\def\gev{{\rm \, Ge\kern-0.1em V}}
\def\tev{{\rm \, Te\kern-0.125em V}}
\def\beq{\begin{equation}}
\def\eeq{\end{equation}}

\def\mchi{m_{\tilde \chi}}
\def\mcha{m_{\chi^{\pm}}}
\def\chiO{\tilde \chi_1^0}

\def\mhd{m_{H_D}}
\def\mhu{m_{H_U}}
\def\m12{m_{1\!/2}}
\def\mz{m_Z}
\def\tb{\tan\beta}

\def\bino{\widetilde B}

\def\bi{\begin{itemize}}
\def\ei{\end{itemize}}
\def\rd{{\rm d}}
\catcode`@=11 \def\citer{\@ifnextchar
  [{\@tempswatrue\@citexr}{\@tempswafalse\@citexr[]}}

\def\@citexr[#1]#2{\if@filesw\immediate\write\@auxout{\string\citation{#2}}\fi
  \def\@citea{}\@cite{\@for\@citeb:=#2\do
    {\@citea\def\@citea{--\penalty\@m}\@ifundefined {b@\@citeb}{{\bf
          ?}\@warning
       {Citation `\@citeb' on page \thepage \space undefined}}%
\hbox{\csname b@\@citeb\endcsname}}}{#1}}
\catcode`@=12

\begin{document}
\begin{titlepage}
\pagestyle{empty}
\baselineskip=21pt
\rightline{MADPH-99-1104}
\rightline{February 1999}
\vskip1.25in
\begin{center}

{\large{\bf Higgsinos in SUSY Models with Gaugino Mass Unification}}
\end{center}
\begin{center}
\vskip 0.5in
{Toby Falk}

{\it Department of Physics, University of Wisconsin, Madison, WI~53706,WI, USA}

\vskip 0.5in
{\bf Abstract}
\end{center}
\baselineskip=18pt \noindent
In the MSSM, the assumptions of a common sfermion mass parameter $m_0$
and a common gaugino mass parameter $\m12$, along with the
requirements from radiative electroweak symmetry breaking, lead to
relatively large values of the Higgs mixing parameter $\mu$, and
consequently to a gaugino-like lightest neutralino $\chiO$.  Lifting
the requirement that the Higgs mass parameters $\mhd$ and $\mhu$ unify
with the sfermion masses is known to allow for smaller $\mu$.  We show
that a $\mu$ parameter sufficiently small to yield a Higgsino-like
neutralino $\chiO$ requires a precise adjustment of the Higgs mass
parameter $\mhu$.  Consequently a gaugino-type neutralino is still
preferred in SUSY models with gaugino mass unification.
\end{titlepage}
\baselineskip=18pt

The composition of the lightest neutralino affects both its
phenomenology and cosmology.  The couplings of the light neutralino
and chargino states, as well as the relationship between the two
lightest neutralino and lightest chargino masses, depend on their
Higgsino content.  Thus the production rates and branching ratios for
charginos and neutralinos, and the expected SUSY signatures associated
with ino production, vary with the ino compositions, and studies of
SUSY searches at future experiments need to include separate analyses
for Higgsino and gaugino-like $\tilde\chi^\pm$ and $\chiO$
\cite{tev2k}.  The cosmology of neutralinos is even more sensitive to
their composition, as a Higgsino-like neutralino tends to have a
significantly lower relic abundance than its gaugino-like counterpart,
and searches for charginos, neutralinos and Higgs bosons at LEP have
all but excluded Higgsinos as giving a dominant contribution to the
energy density of the universe \cite{efgos}.  Further, the interaction
rates of relic neutralinos with nucleii depend sensitively on the
neutralino composition.  It is therefore interesting to know what
restrictions one may place on Higgsino content of the neutralino.  In
particular, we consider the standard lore, that relaxing the scalar
mass unification condition to allow for non-universal Higgs mass
parameters $\mhu$ and $\mhd$ permits the neutralino $\chiO$ to be
either Higgsino-like or gaugino-like.  We will show that while it is
true that the Higgs mixing mass $\mu$ may be chosen small enough to
provide a Higgsino-like LSP, it is at the price of choosing a very
particular narrow range in $\mhu$.

It is well known that in models with both gaugino and scalar 
mass unification, the lightest neutralino tends to be a gaugino, and
in particular a bino.    $|\mu|$ is fixed at the
electroweak scale by the Higgs potential minimization condition, which 
at tree level reads
\begin{equation}
  \label{eq:min}
  |\mu|^2={\mhd^2-\mhu^2 \tan^2\beta\over\tan^2\beta-1} - {\mz^2\over2}.
\end{equation}
Since $\mhd(M_X)$ and $\mhu(M_X)$ are fixed by the scalar mass unification condition, their
low-energy values are determined by their RGE evolution to the
electroweak scale.   The dominant parts of the one-loop RGEs for  $\mhd$ and $\mhu$ are
given by
\begin{eqnarray}
  \label{eq:rge}
  {\rd\mhd^2\over\rd t}&=&{1\over 8\pi^2}\left(-3 g_2^2 M_2^2 - g_1^2 M_1^2 
  + 3 h_b^2 (m_{\tilde Q_L}^2 + m_{\tilde b_R}^2 + \mhd^2 +
  A_b^2) - {1\over2}g_1^2 S\right)\nonumber\\
  {\rd\mhu^2\over\rd t}&=&{1\over 8\pi^2}\left(-3 g_2^2 M_2^2 - g_1^2 M_1^2 
  + 3 h_t^2 (m_{\tilde Q_L}^2 + m_{\tilde t_R}^2 + \mhu^2 +
  A_t^2) + {1\over2}g_1^2 S\right),
\end{eqnarray}
where
\begin{eqnarray}
  \label{eq:dterm}
  S=\mhu^2-\mhd^2 + \sum_{i=1}^3m_Q^2-m_L^2-2 m_u^2+m_d^2+m_e^2.
\end{eqnarray}
The terms proportional to $S$ are the D-term contributions
\cite{ikkt,mv} to the running of the scalar mass$^2$ parameters.  In
mSUGRA, $S$ vanishes at the unification scale, and as $S=0$ is a fixed
point of the RG equations\cite{ikkt}, it remains 0 at all scales.
Symmetries force $S$ to vanish in many models of interest, and
consequently the D-term contribution to the scalar mass$^2$ beta
functions do not appear in many compilations of the SUSY 1-loop RGEs.
Since we will be breaking scalar mass unification, we need to keep the
$S$ terms on the right-hand side of (\ref{eq:rge}).  

\begin{figure}[thb]
\begin{center}
\vspace*{-0.2in} 
\hspace*{-0.2in}
\epsfig{file=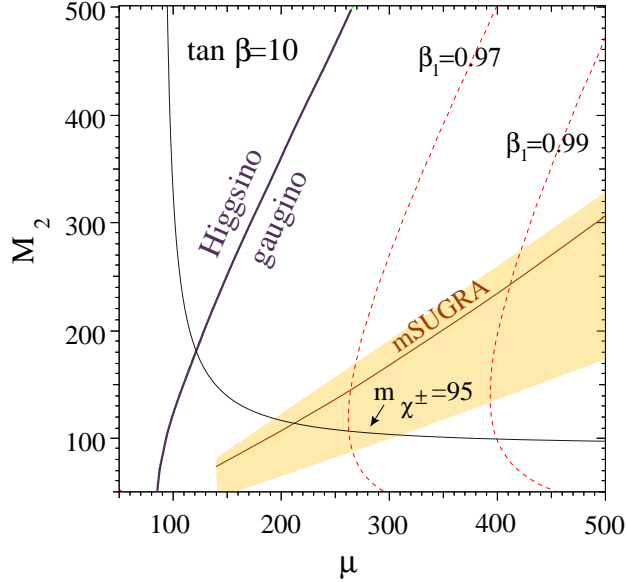,height=3in}
\end{center}
\vspace{-0.2in} 
\caption{{\it 
    The gaugino and Higgsino regions in the $\{\mu,M_2\}$ plane.  The
    dashed curves are bino purity contours.  Also shown are the
    mSUGRA solution for $\mu(M_2)$ for $\tb=10, m_0=100\gev$, along
    with the current LEP lower bound of 95 $\gev$ on the mass of the
    lightest chargino. }} {
  \label{fig:purity} } 
\end{figure}

It is the positive terms proportional to the top Yukawa coupling which
drive $\mhu^2$ negative and allow (\ref{eq:min}) to be satisfied in
the standard scenario.  Since the squark mass$^2$ parameters receive a
large contribution in their evolution proportional to the square of
the gluino mass, and since $A_t$ approaches its quasi-fixed-point
value proportional to $\m12$ (as does $A_b$ at large $\tb$, where
$h_b$ is significant), the right-hand side of (\ref{eq:rge}), and
hence the changes in $\mhd^2$ and $\mhu^2$ as they are evolved to the
electroweak scale, scale with the gaugino masses$^2$.  Neglecting the
small $\mz^2$ term, the final value of $\mu$ given by (\ref{eq:min})
is then simply proportional to $\m12$.  The inclusion of radiative
corrections to (\ref{eq:min}) do not significantly alter this result,
although they can move the regions where scaling is violated.  An
example of the scaling between $\mu$ and the gaugino mass is given in
Fig.~\ref{fig:purity}, where the minimal supergravity (mSUGRA)
solution for $\mu$ is shown in the $\{M_2,\mu\}$ plane at fixed
$m_0=100\gev$, with $A_0=0$.   At intermediate and large $\tan\beta$, 
$A_t$ can of course stray far from the quasi-fixed point when $|A_0|\gg\m12$.
Large $|A_t|$ increases the effect of the top Yukawa coupling in (\ref{eq:rge})
and can subsequently increase $|\mu|$.  This effect  is limited by the fact
that for $A_0$ too large, the right stop mass$^2$ parameter is driven negative; however,
a significant fractional increase in $|\mu|$ can still be produced, particularly for
large $m_0$.   By contrast, it is difficult to reduce $|\mu|$ by any significant amount,
as the magnitude of $A_t$ is $\ga{\cal O}(\m12)$ over most of its
evolution, even if it doesn't closely approach the quasi-fixed point.  In Fig.~\ref{fig:purity},
$|\mu|$ takes its minimum value for $A_0\sim -2\m12$ \footnote{In our
  sign conventions, the mixing term in the stop mass matrix is
  $-m_t(A_t+\mu\cot\beta)$.}, where it is reduced  from
its $A_0=0$ value by $\sim$10\%.  The light shaded band shows the possible
range of $\mu$, varying over allowed $A_0$.

In general, the neutralinos are linear combinations of the neutral gauginos and
Higgsinos,
\begin{equation}
\chi_i = \beta_i  {\tilde B} + \alpha _i{\tilde W} + \gamma_i{\tilde
H}_1 + \delta _i{\tilde H}_2, \hspace{0.3in}  i=1,\ldots, 4
\label{neut}
\end{equation}
In this notation, the gaugino purity of a neutralino
$\chi_i$ is defined to be $p_i=\sqrt{{\alpha_i}^2 + {\beta_i}^2}$, and its Higgsino
purity $\sqrt{1-p_i^2}=\sqrt{{\gamma_i}^2 + {\delta_i}^2}$.  
In the $({\tilde B}, {\tilde W}^3, {{\tilde H}^0}_1,{{\tilde
    H}^0}_2 )$ basis, the neutralino mass matrix takes the form
\begin{equation}
\left( \begin{array}{cccc}
M_1 & 0 & {-M_Z s_\theta c_\beta} &  {M_Z s_\theta s_\beta} \\ 0 & 
M_2 & {M_Z c_\theta c_\beta} & {-M_Z c_\theta
s_\beta}
\\ {-M_Z s_\theta c_\beta} & {M_Z c_\theta c_\beta} & 0
& -\mu
\\ {M_Z s_\theta s_\beta} & {-M_Z c_\theta s_\beta} &
-\mu & 0 
\end{array} \right) ,
\label{mm}
\vspace{0.1cm}
\end{equation}
where $s_\theta$ ($c_\theta$) $=\sin\theta_W$ ($\cos\theta_W$),
$s_\beta$ ($c_\beta$) $=\sin\beta$ ($\cos\beta$), and where gaugino
mass unification implies $M_1= 5/3 \tan^2\theta_W M_2\approx
0.4\,\m12$.  In the limit $|\mu|\gg M_i$, the lightest neutralino is
gaugino-like, specifically a $\bino$, with mass $\mchi\approx M_1$.
Only if $|\mu|\la M_1$ will the lightest neutralino be Higgsino-like,
with mass $\mchi\approx|\mu|$.  Superimposed in Fig.~\ref{fig:purity}
are contours of constant 99\% and 97\% bino purity $|\beta_1|$, showing that the
lightest neutralino in mSUGRA tends to be quite pure gaugino.  The
dark thick contour, which separates the gaugino-like from
Higgsino-like neutralinos and corresponds to $p_1= \sqrt{1-p_1^2}=1/\sqrt{2}$, lies
well to the left of the mSUGRA contour.  Thus the prejudice towards a
bino as the lightest neutralino.

The current LEP lower limit of $95\gev$ on the mass of the lightest
chargino is shown as the thin solid line in Fig.~\ref{fig:purity}.
The displayed chargino mass includes the full 1-loop corrections
\cite{pp}, which give it only a very mild dependence on $m_0$ (here
taken to be $100\gev$).  The chargino mass constraint provides an absolute lower
bound on $\mu$.  Further, the intersection of the chargino bound with
the Higgsino/gaugino threshold lies at $M_2\sim 180\gev$, implying
that a Higgsino-like neutralino is excluded by the chargino limits
alone for $\m12< 220\gev$, for this value of $\tb$ and sign of $\mu$.
Similar corresponding bounds apply for other $\tb$ and for $\mu<0$.

\begin{figure}[thb]
\begin{center}
\vspace*{-1.4in} 
\hspace*{-0.2in}
\epsfig{file=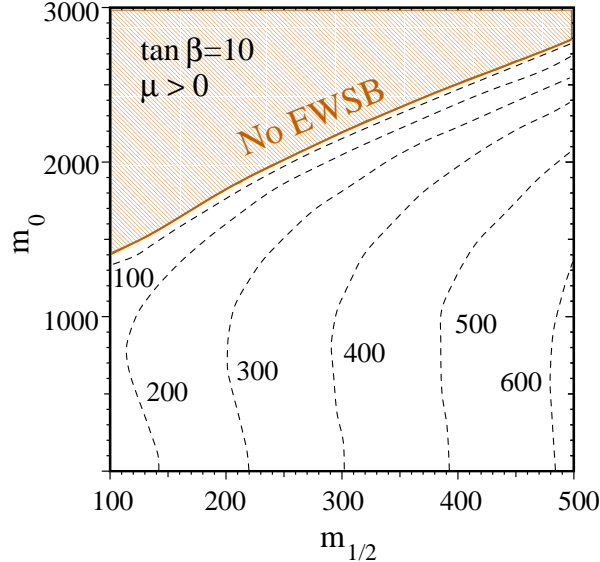,height=5in}
\end{center}
\vspace{-1.0in} 
\caption{{\it Contours of constant $\mu$, for $\tb=10, A_0=0$. The
    upper shaded region does not admit correct electroweak symmetry breaking.}} {
  \label{fig:noscaling} } 
\end{figure}

For sufficiently large $m_0 \gg \m12$, the scalar mass contributions
to the squark and Higgs mass parameters in (\ref{eq:min}) and
(\ref{eq:rge}) remain important, and scaling can break down.  This is
shown in Fig.~\ref{fig:noscaling}, where contours of constant $\mu$
are plotted in the $\{\m12,m_0\}$ plane for $\tb=10$; note the
expanded scale of the $m_0$ axis compared to the $\m12$ axis.  In all
the figures we display, we use two loop RGEs\cite{mv} to evolve the
dimensionless couplings and the gaugino masses, and one loop
RGEs\cite{ikkt,dn} for the other soft masses, and we include one-loop
SUSY corrections to $\mu$\cite{bbo} and to the top and bottom
masses\cite{pbmz} .  For large $\m12/m_0$, $(\mu/\m12)\sim1.25-1.35$.
Note that the contours in Fig.~\ref{fig:noscaling} are almost vertical
for $m_0 \le \m12$, but deviate from vertical for $m_0\gg \m12$.  For
sufficiently large $m_0$, $|\mu|^2$ determined by (\ref{eq:min})
becomes negative, and the vacuum does not exhibit correct electroweak
symmetry breaking.  Since large cancellations are required to allow
$\mu=0$, the location of the line at which this happens is sensitive
to the radiative corrections to (\ref{eq:min}), as well as to the loop
order at which the masses and couplings are run.  However, it always
is positioned at large $m_0\gg\m12$.  This region is marked by
dark shading.

In Fig.~\ref{fig:noscaling}, there are small regions at large $m_0$,
near to where $|\mu|^2$ vanishes in (\ref{eq:min}), where
$\mu$ is much smaller than that given by the na\"{\i}ve scaling
relation.  However, the size of the region with $|\mu|<0.4\m12$ is
tiny, and it is further reduced by the LEP
chargino mass constraint.  If we want a Higgsino LSP, hence small
$|\mu|$, we must either live in the tiny fringe areas at large $m_0$,
and content ourselves with a very heavy scalar spectrum, or we must
break either the scalar or gaugino mass unification condition.
Clearly, if $\mhd$ and $\mhu$ are free parameters in (\ref{eq:min}),
$\mu$ can take any value, and it is part of the standard lore that
breaking scalar mass unification for the Higgs mass parameters allows
for a completely general set of Higgs sector masses and mixings.  It
is this standard claim we wish to examine in this paper.

\begin{figure}[thbp]
\vspace{-1.8in}
\begin{minipage}{5.5cm}
\hspace*{-0.3in}
\epsfig{file=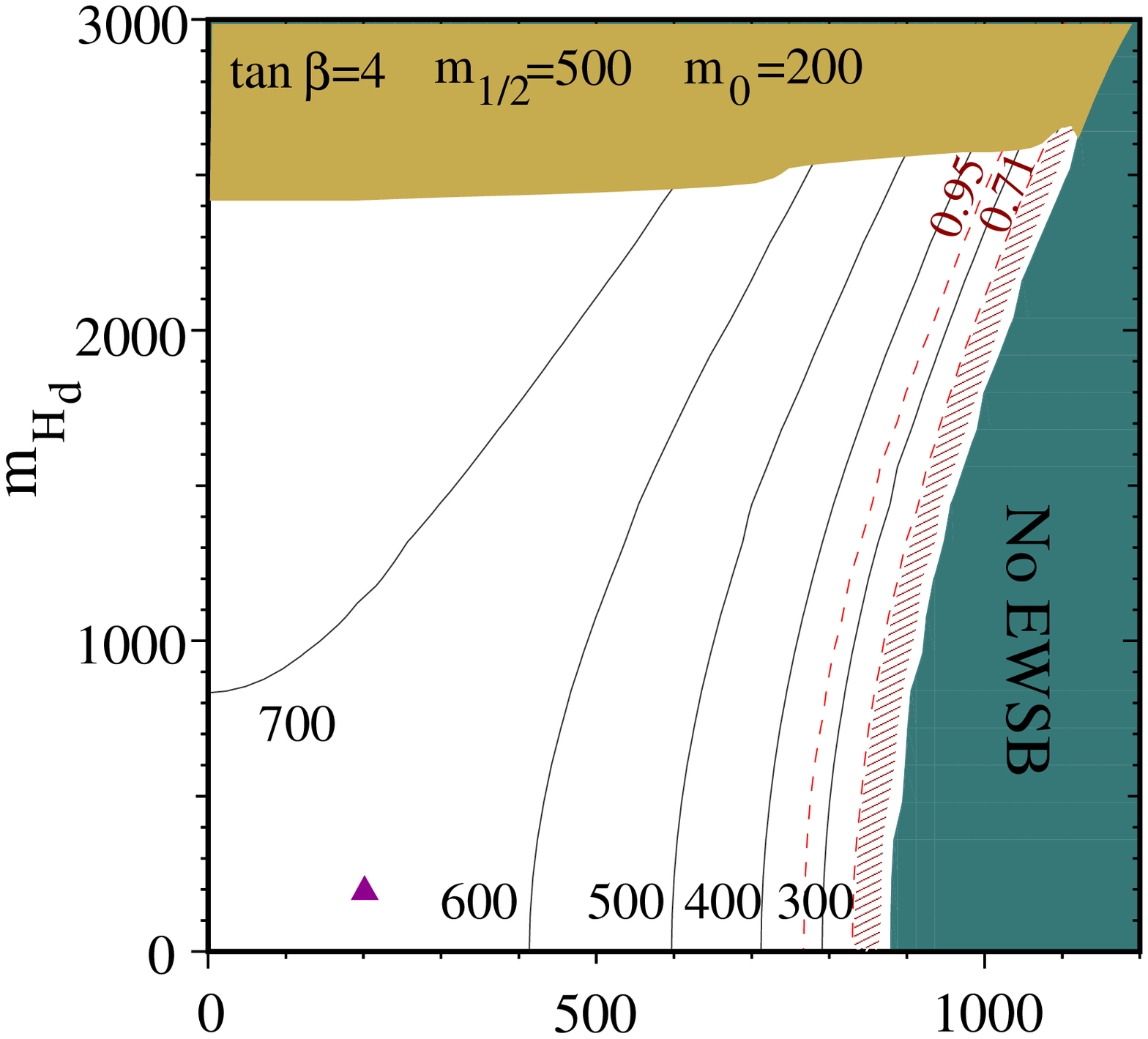,height=5in}     
\end{minipage}
\hspace*{0.5in}
\begin{minipage}{5.5cm}
\epsfig{file=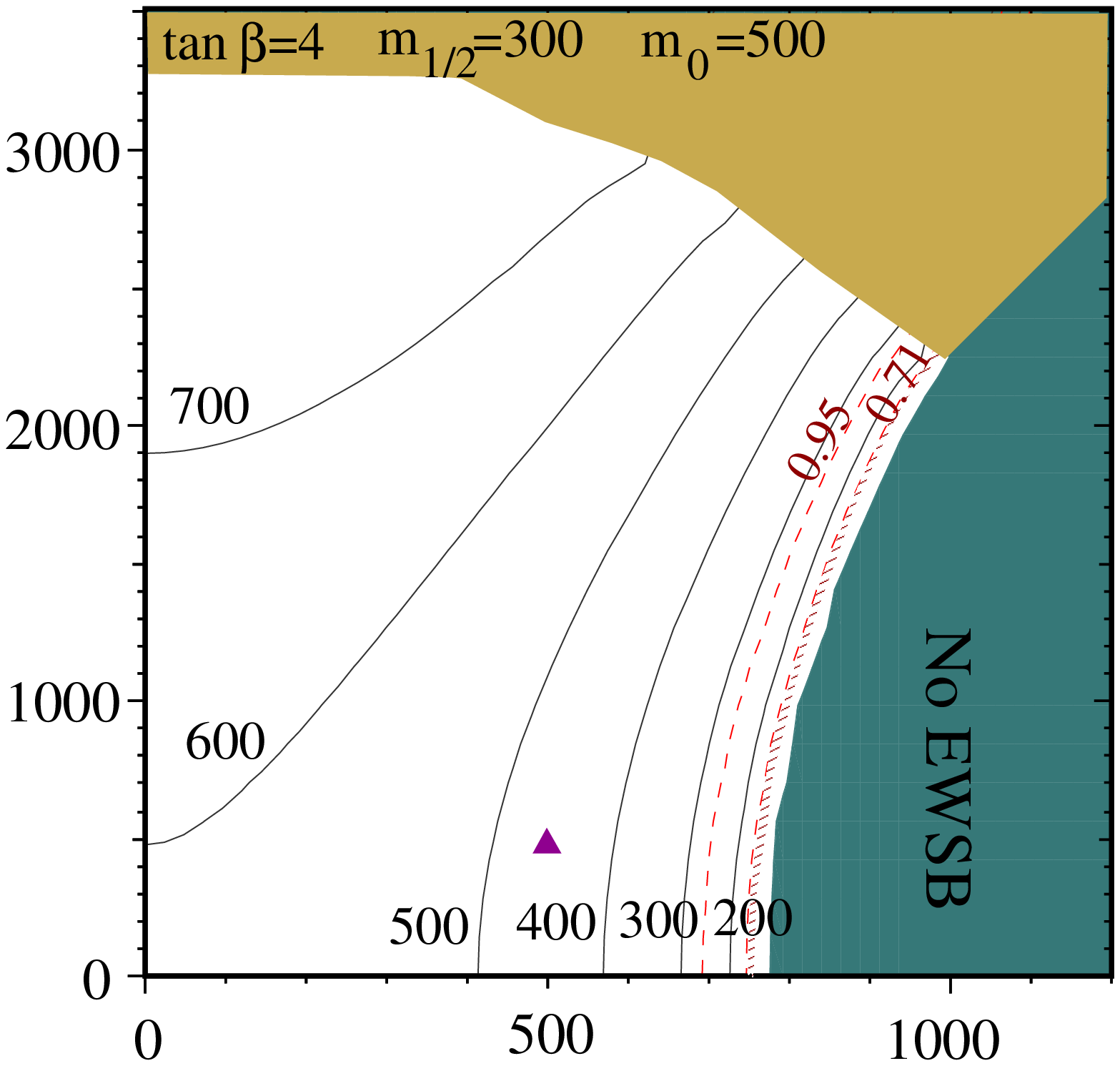,height=5in} 
\end{minipage}\hfill
\vspace{-2.3in}
\begin{minipage}{5.5cm}
\hspace*{-0.3in}
\epsfig{file=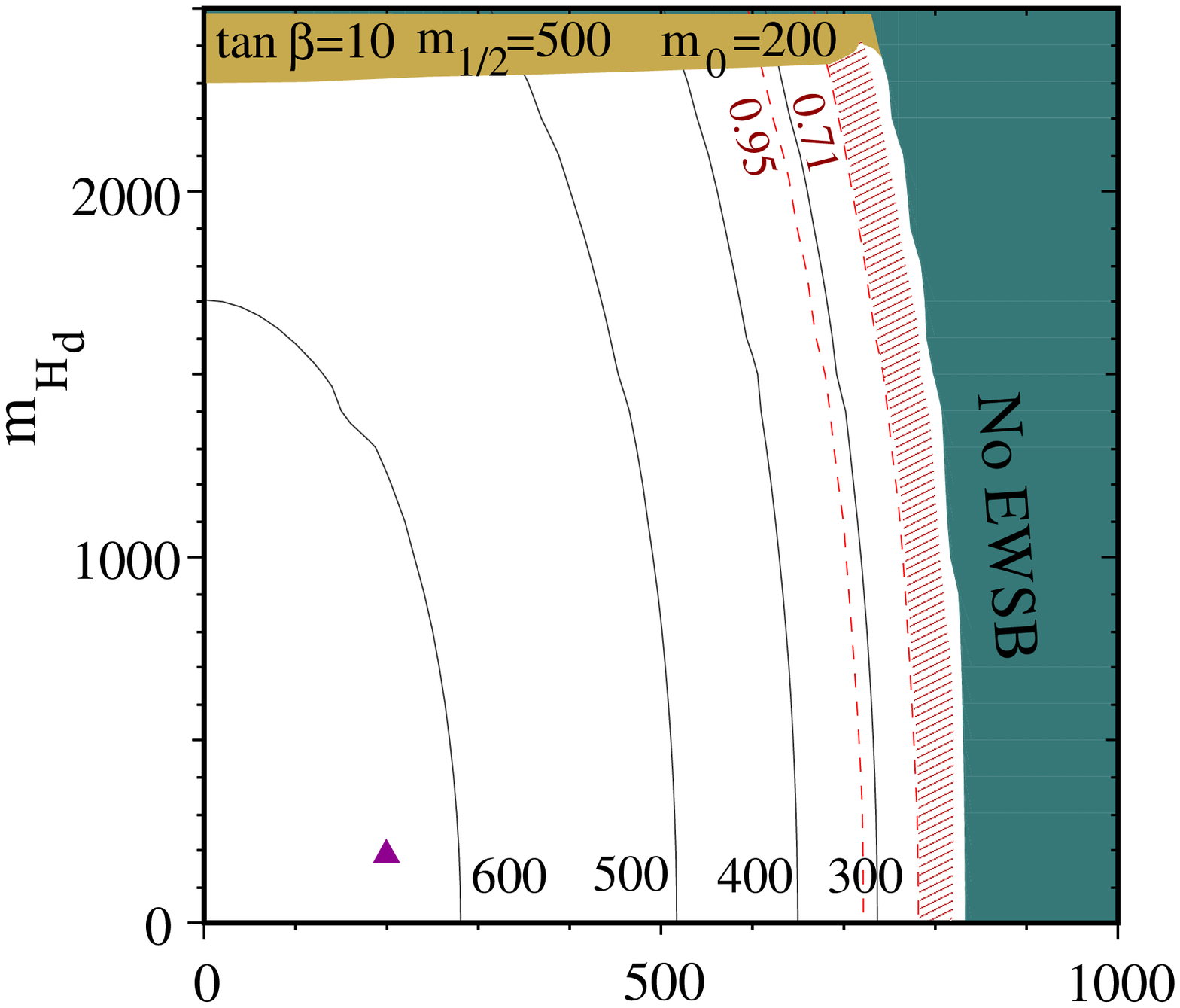,height=5in}
\end{minipage}
\hspace*{0.5in}
\begin{minipage}{5.5cm}
\epsfig{file=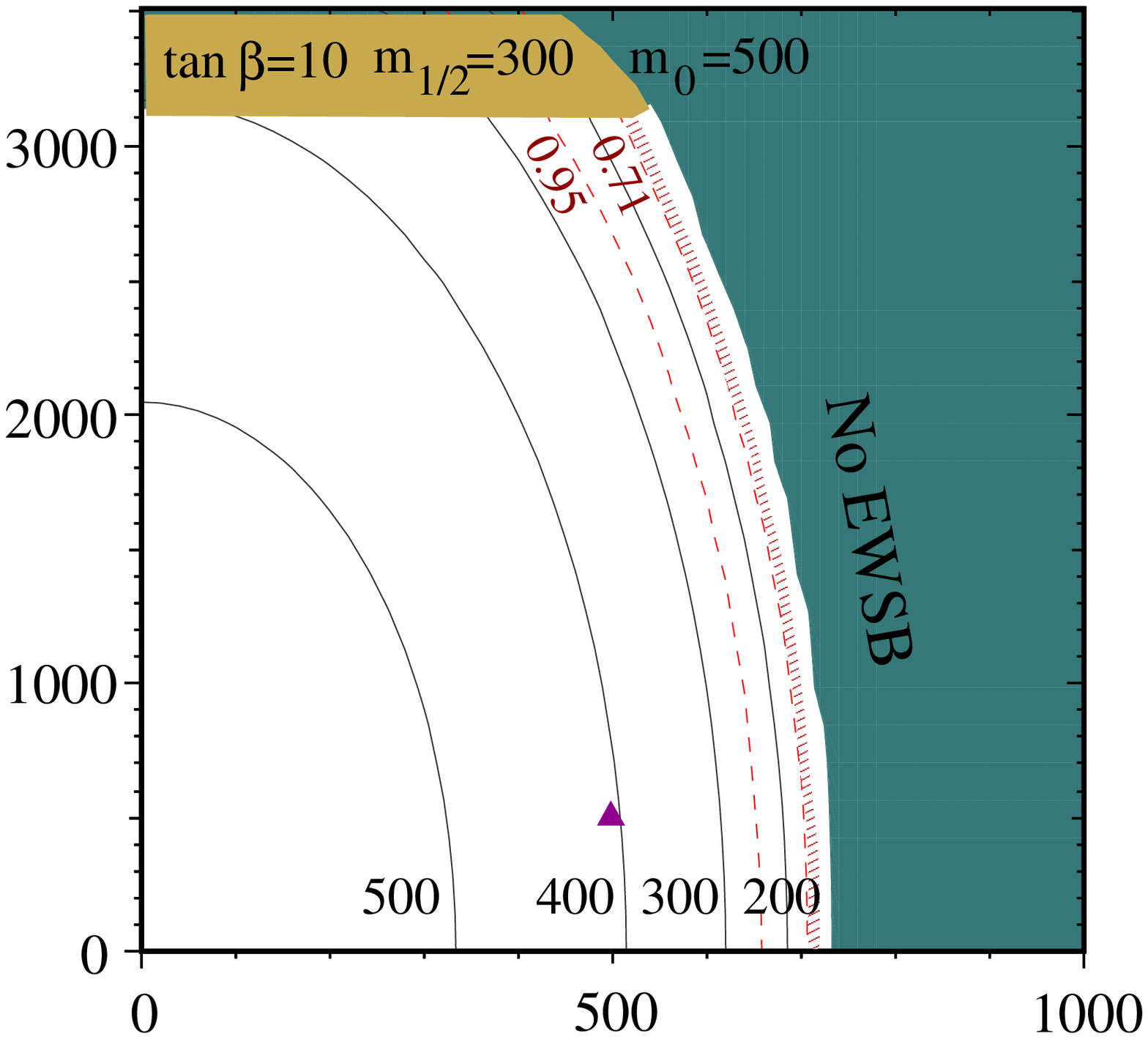,height=5in} 
\end{minipage}\hfill
\vspace{-2.3in}
\begin{minipage}{5.5cm}
\hspace*{-0.3in}
\epsfig{file=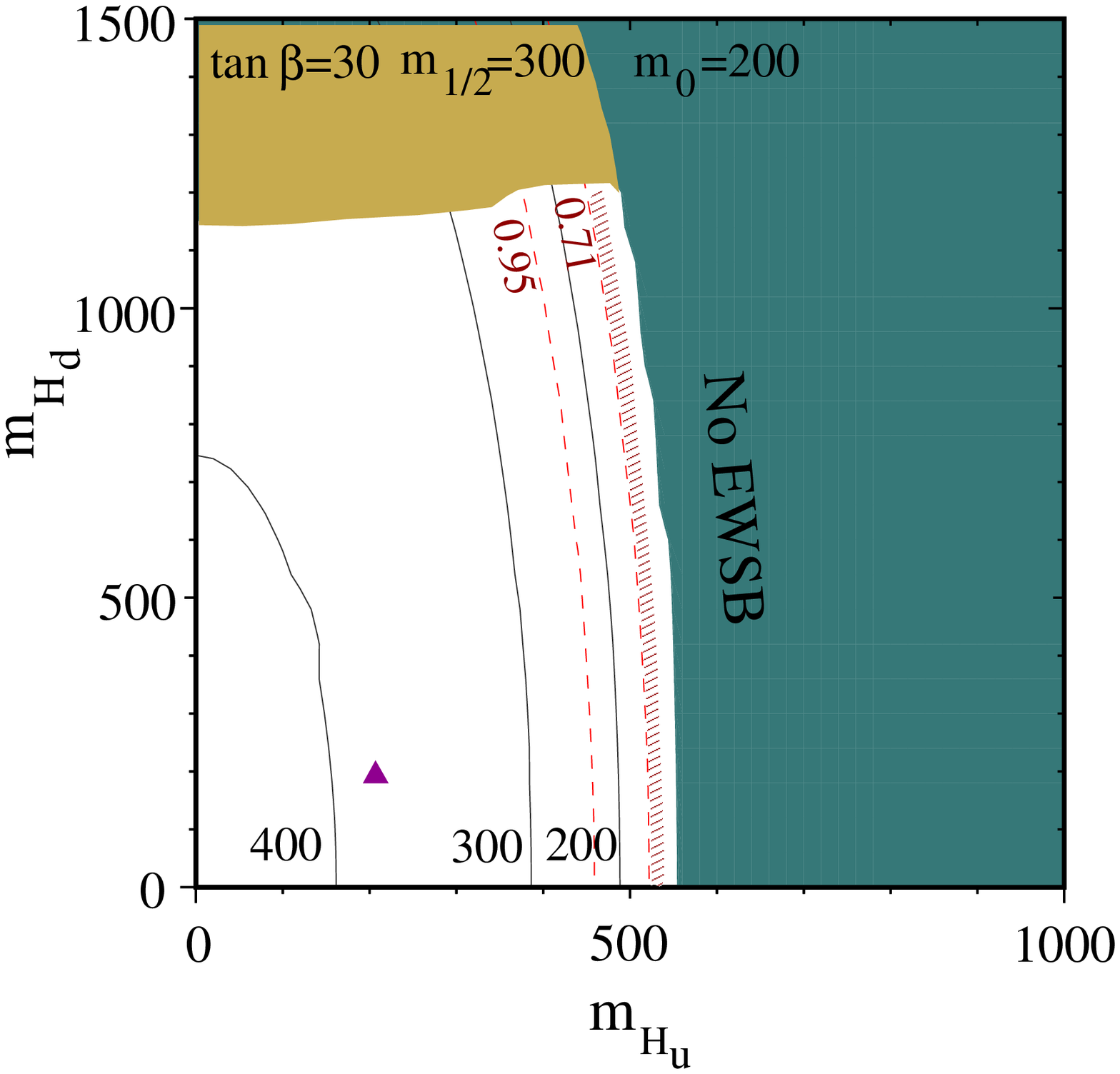,height=5in}     
\end{minipage}
\hspace*{0.5in}
\begin{minipage}{5.5cm}
\epsfig{file=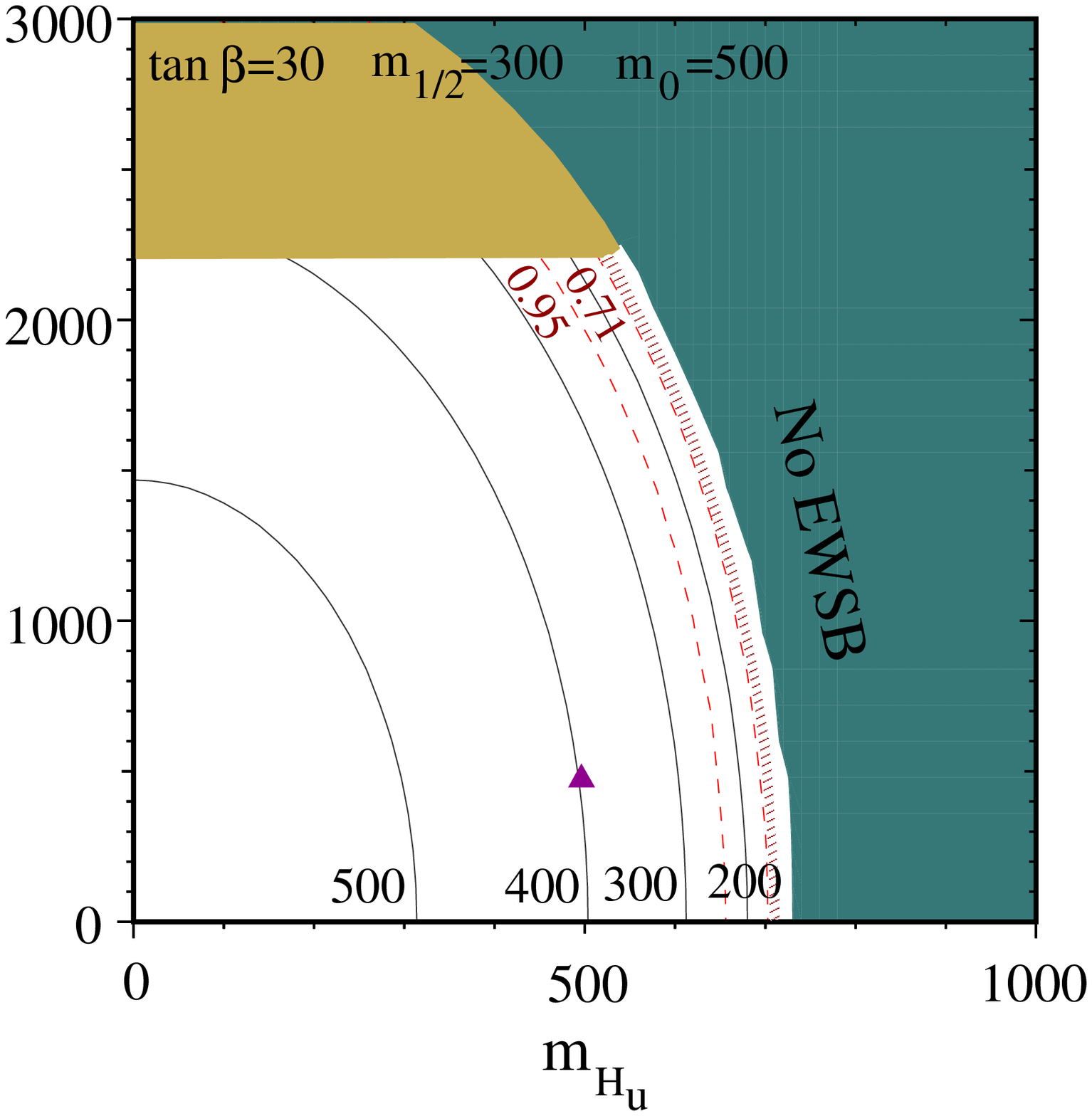,height=5in} 
\end{minipage}\hfill
\vspace{-0.9in} 
\caption{{\it Contours of constant $\mu$ (solid lines) in the $\{\mhu,\mhd\}$
    plane.  Dashed lines are contours of gaugino purity $p_1=0.95$ and
    $p_1=0.71$.  The hashed area is the part of the Higgsino region
    allowed by the LEP chargino mass constraint.  The shaded areas
    are excluded either because they do not allow correct electroweak
    symmetry breaking (dark) or they yield a tachyonic sfermion
    (light).  The mSUGRA point is marked by a triangle.}} {
  \label{fig:mumd} } \vspace{0.3in}
\end{figure}

Accordingly, we now relax the mSUGRA unification constraints assumed
in Fig.~\ref{fig:noscaling} and allow $\mhd$ and $\mhu$ to vary
freely.  Fig.~\ref{fig:mumd} displays contours of constant $\mu$ in
the $\{\mhu,\mhd\}$ plane, for two values of fixed $\m12$ and sfermion
mass $m_0$, and for three values of $\tb$.  Plots for $\mu<0$ are very
similar.  The shaded regions have either a tachyonic stop or stau
(typically for large $\mhd$) or do not permit electroweak symmetry
breaking (for large $\mhu$).  The mSUGRA point is marked by a
triangle.  Gaugino purities of 0.95 and 0.71 ( the purity value at
which the lightest neutralino is half gaugino and half Higgsino) are
denoted by dashed contours.  The Higgsino region is a thin little
strip, tucked up against the excluded shaded region.  The hatched area
is the part of the Higgsino region which satisfies the chargino mass
constraints\footnote{The chargino bounds are weakened when the
  chargino is sufficiently degenerate with the neutralino, as in the
  pure Higgsino limit, for large enough $|\mu|$.  We verify that
  $\mcha-\mchi > 10\gev$, so that the chargino limit is not reduced.}.
This graphically demonstrates the careful adjustments required to
achieve a Higgsino-like lightest neutralino.

\begin{figure}[thbp]
\vspace{-1.5in}
\begin{minipage}{6.0cm}
\hspace*{-0.2in}
\epsfig{file=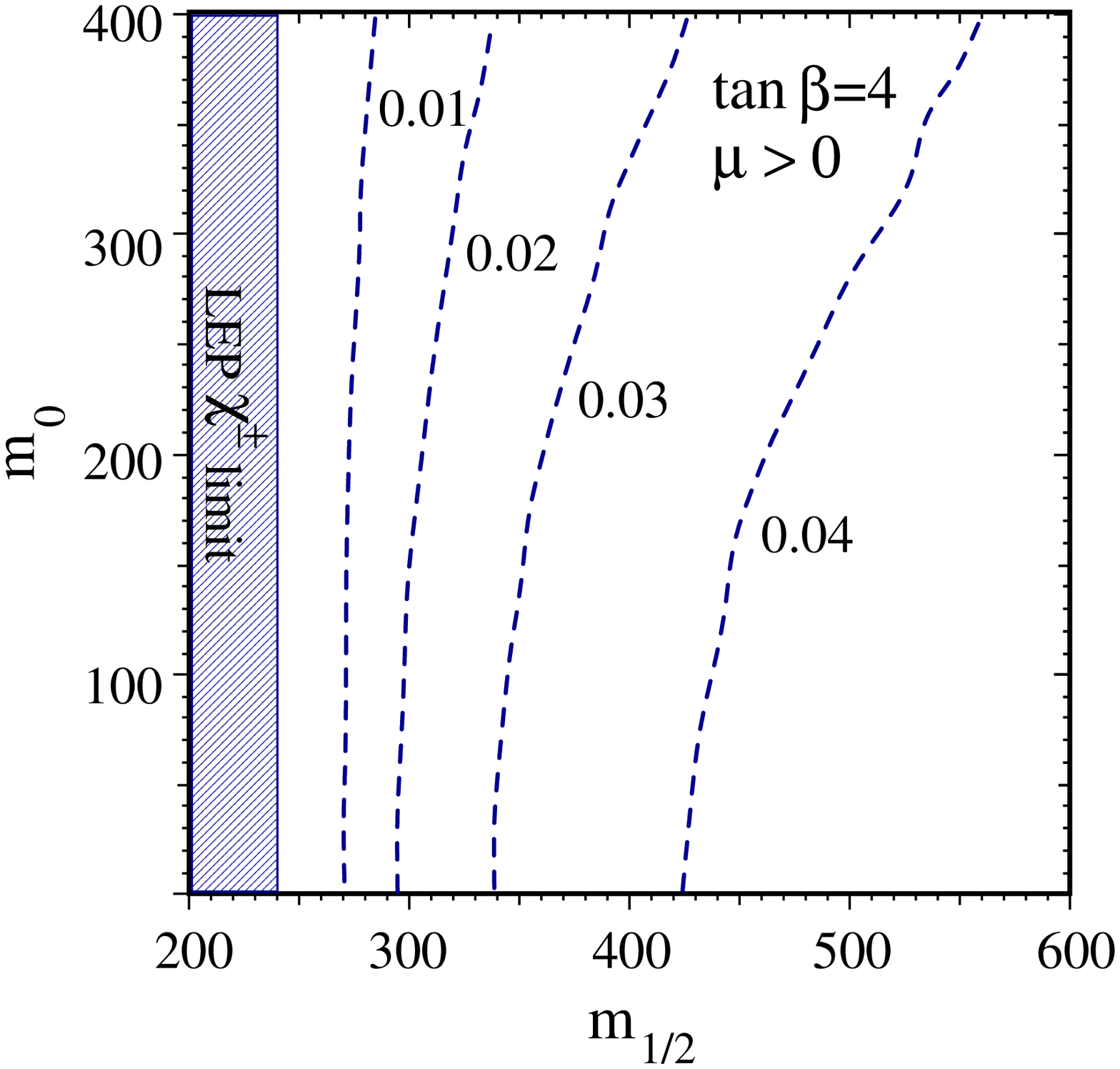,height=5.5in}
\end{minipage}
\hspace*{0.5in}
\begin{minipage}{6.0cm}
\epsfig{file=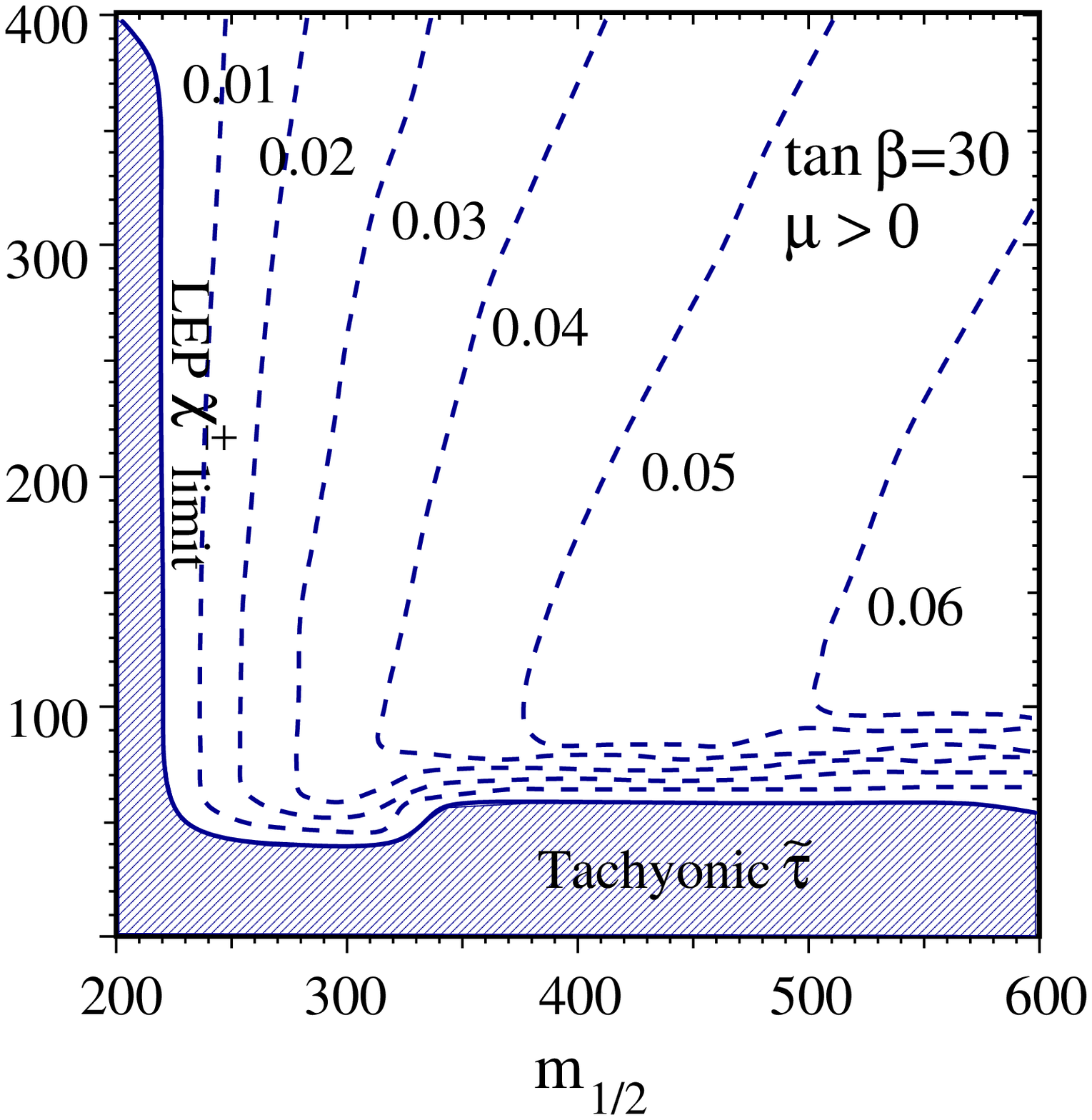,height=5.5in} 
\end{minipage}\hfill
\vspace{-2.0in}
\begin{minipage}{6.0cm}
\hspace*{1.0in}
\epsfig{file=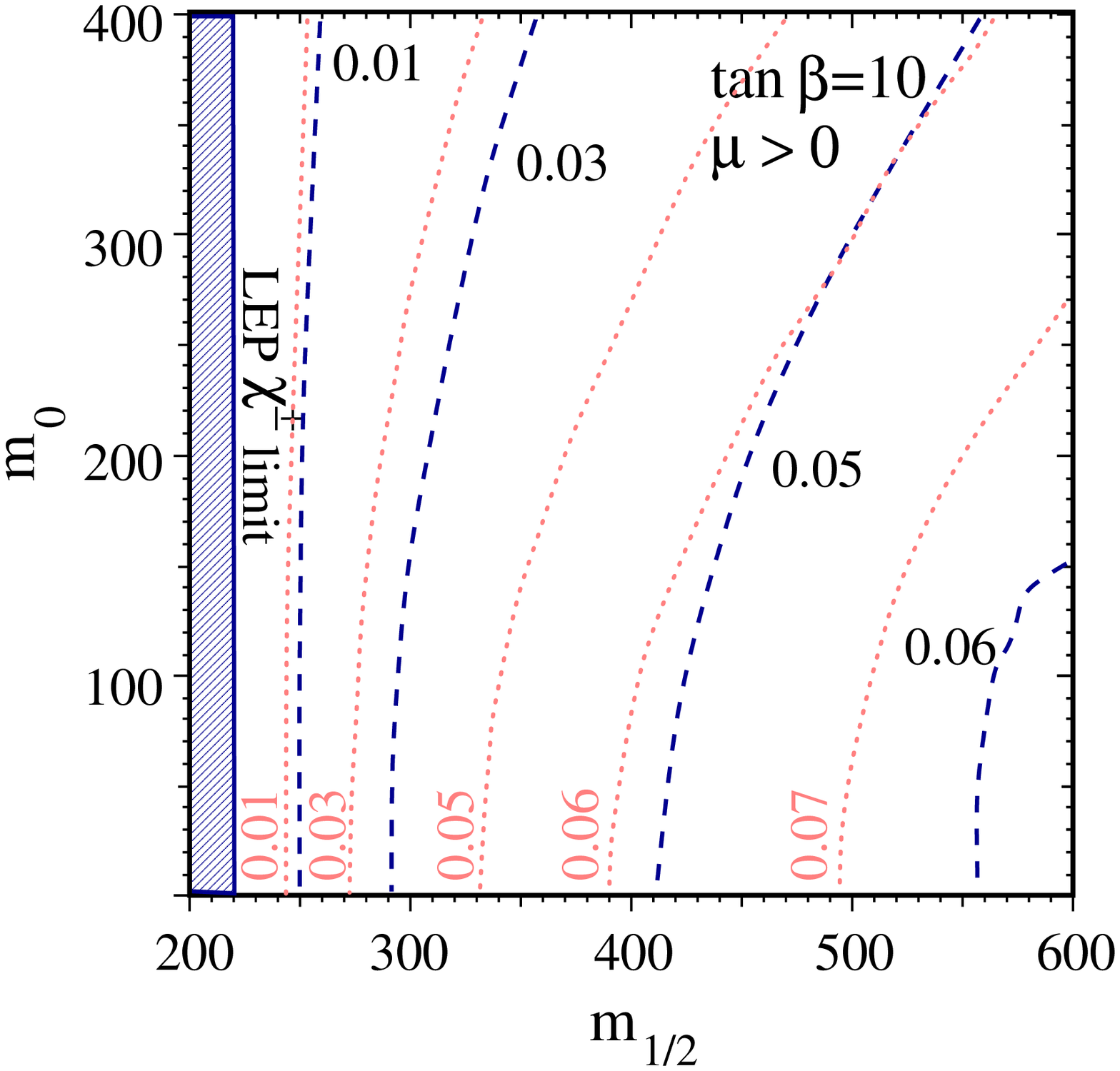,height=5.5in}     
\end{minipage}\hfill
\vspace{-0.5in} 
\caption{{\it The fraction of the allowed $\mhu$ parameter space which 
    yields a Higgsino-type lightest neutralino, for $\tb=\{4,10,30\}$
    and $\mu>0$.  The fractional areas for
    $\mu<0$ are similar.  In the shaded areas, the entire Higgsino
    region is excluded either by the LEP chargino bound or by the
    presence of tachyonic sfermions.  We've fixed $\mhd=\mhu$ and
    $A_0=0$.  The light dotted contours for $\tb=10$ show the effect
    of taking $A_0=-2\m12$. }} {
  \label{fig:area} } \vspace{0.3in}
\end{figure}

The position of the Higgsino region varies with $m_0$ and $\m12$, but
to a good approximation, for $\mhd=\mhu$, the top end of the allowed
Higgsino region lies at
\begin{equation}
  \label{eq:mhu}
  \mhu^2\approx 2.6\, \m12^2 + 1.0\, m_0^2, 
\end{equation}
for $4\la\tb\la30$.  We've taken $A_0=0$ in Fig.~\ref{fig:mumd}; while
the position of the region does vary some with $A_0$, the size of the
hatched region remains small.  Since $\mu$ is more sensitive to $\mhu$
than $\mhd$ in the allowed Higgsino region, we now for simplicity fix
$\mhd=\mhu$, and we plot in the $\{\m12,m_0\}$ plane the percentage of
the experimentally allowed $\mhu$ parameter space which contains a
Higgsino-like $\chiO$; i.e. we display in Fig.~\ref{fig:area} contours
of constant $({\mhu}^{\rm max} -{\mhu}^{\rm min})/{\mhu}^{\rm max}$.
In the shaded areas, the Higgsino regions are either entirely excluded
by the LEP chargino constraints or yield a tachyonic stau. Typically
less than 5\% of the experimentally available range of $\mhu$ yields a
Higgsino-like lightest neutralino.  The dashed contours are for
$A_0=0$; taking $A_0=-2\m12$ yields the dotted contours of the
$\tb=10$ panel of Fig.~\ref{fig:area}, while taking $A_0>0$ makes the
allowed Higgsino regions smaller than for $A_0=0$.  The areas are insensitive to the sign
of $\mu$.  We note also that the entire allowed areas also satisfy the
current LEP2 Higgs mass constraints.  

We parenthetically note that the particular choice $\mhd=\mhu$ is
convenient, because it allows the D-term contributions to the running
of the scalar mass$^2$ parameters to vanish at one loop.  For large
scalar mass non-universalities, the effect of the $S$ term in the RGEs
can be substantial, and the direction of the effect depends on the
pattern of mass differences.  This can lead to tachyonic sfermions in
some cases where, in the universal case, the masses are well behaved.
When $\mhu=\mhd$, and the other scalar masses are universal, $S$
vanishes, and this complication is avoided.  In practice, taking
$\mhd=m_0$ instead of $\mhd=\mhu$ produces only a small downward shift
in the contours in Fig.~\ref{fig:area}, but some of the allowed
Higgsino regions in Fig.~\ref{fig:area} at small $m_0$ now contain a
tachyonic stau.  

The Higgsino regions typically lie at $\mhu > m_0$, and for $\m12\gg
m_0$, (\ref{eq:mhu}) gives $\mhu\gg m_0$.  This is in contrast to
expectations for the pattern of scalar mass non-universality coming,
for example, from evolution of the soft masses from $M_P$ to $M_{\rm
  GUT}$ in models where the soft masses unify at $M_P$ rather than
$M_{\rm GUT}$ \cite{popo}.  These effects in fact tend to reduce
$\mhu$ from its unified value, rather than enlarge it, and the
concomitant generic increase in $|\mu|$ and consequent decrease in the
neutralino Higgsino content are documented in \cite{popo}.  However,
once universality is violated at the GUT scale, D-terms associated
with the reduction in rank of the gauge group to the SM can provide
additional non-universal contributions to the soft masses, and these
can in principle yield a suffiiciently large $\mhu$ \cite{popo,mop}.
Rather than study the effect of non-universality patterns from
specific models, it is common (see e.g. \cite{op1,ccn,cos}) to
parameterize the GUT scale non-universalities by introducing factors
$\delta_i$, so that, e.g.  $\mhd^2=(1+\delta_{H_d})m_0^2$ and
$\mhu^2=(1+\delta_{H_u})m_0^2$, and to allow the range $-1\le
\delta_{H_d},\delta_{H_u}\le 1$, in accord with expectations from
\cite{popo}.  This spread of $\delta_i$ is chosen to comfortably
encompass the expected deviations from universality from GUT scale
effects.  Comparing to (\ref{eq:mhu}), we see that this permits a
Higgsino-type neutralino for $m_0\ga\m12$.

It is useful to compare our results with previous discussions of
fine-tuning in mSUGRA and the MSSM \citer{tuning,kk}.  It has been
observed \cite{tuning} that mSUGRA itself exhibits fine-tuning, in
part due to the condition (\ref{eq:min}), which requires a large
cancellation between the soft mass terms when $\m12$ is much larger
than $\mz$.  It is common to introduce the fine-tuning parameters
$\Delta_a$\cite{tuning}, which describe the sensitivity of the
electroweak scale to variations of each parameter $a$ in the model,
and where\footnote{An alternate definition, where $\mz^2$ is replaced
  by $\mz$, is a factor of 2 smaller. See \cite{anderson} for an
  alternate tuning measure which compares the parameter sensitivity to
  an ``average'' sensitivity.  See also \cite{ccn} and \cite{wright}
  for other tuning measures.}
\begin{equation}
  \label{eq:delta}
  \Delta_a=\left| {a\over \mz^2}{\partial \mz^2\over\partial a}\right|.
\end{equation}
The largest $\Delta_a$ gives the degree of fine-tuning in the model,
for a fixed parameter set.  Minimal SUGRA is typically characterized
by large $\Delta_\mu$, since small changes in $\mu$ upset the delicate
cancellations in (\ref{eq:min}) and produce large (fractional) changes
in $\mz$.  Relaxing the scalar mass unification
condition\citer{op1,kk} for $\mhd$ and $\mhu$ can
somewhat ameliorate the fine-tuning problem, precisely because their
low-energy values can be adjusted to yield a smaller value for $\mu$
in (\ref{eq:min}).  These studies are germane to our problem, as small
$\mu$ is required for a Higgsino-like LSP.  However, no one to date
has specifically addressed the issue of the conditions required to
obtain a Higgsino-like LSP.

In fact, $\Delta_\mu$ is much smaller in the Higgsino regions in
Figs.~\ref{fig:noscaling} and \ref{fig:mumd} than in the generic
regions of Fig.~\ref{fig:noscaling}.  However, the largest of the
sensitivity parameters always remains large.  In
Fig.~\ref{fig:noscaling}, $\Delta_{m_0} > 100$ in the region of
interest, and in Fig.~\ref{fig:mumd}, $\Delta_{\mhu}$ is typically $>
80$.  Furthermore, the size of the allowed Higgsino regions are
reduced, as discussed above, by the lower bound on the mass of the
lightest chargino, which forbids the smallest values of $|\mu|$.  This
fact is not reflected in the sensitivity parameter $\Delta$, which is
essentially a local function of the MSSM parameters.  At very large
$\tan\beta\sim50$, one can find Higgsino-like neutralinos with lower
values of $\Delta$ \cite{op1}.  However, one still finds similar
percentage allowed areas to those in Fig.~\ref{fig:area}.  A full
sensitivity analysis which includes all the $\Delta_a$ is beyond the
scope of this paper.

By contrast, a Higgsino-like  lightest neutralino is  more plausible
in the absence of gaugino mass unification.
For starters, breaking the gaugino mass unification
condition alters the RGE evolution of the squarks, and hence
of $\mhd$ and $\mhu$, and thus admits smaller $\mu$.  As recently
emphasized \cite{kk}, since  the dominant terms in (\ref{eq:rge})
driving $\mhu^2$ negative scale with the squark masses$^2$, which 
in turn receive large contributions $\sim M_3^2$,
a reduction of $M_3$ below its unification value 
can significantly reduce the value of $|\mu|$ inferred from
(\ref{eq:min})  (and hence the tuning associated with the $\mu$
parameter), particularly at low $\tb$.  Alternatively,  putting $M_1$ 
above its unification value increases the size of the Higgsino region in the
$\{\mu,M_2\}$ plane by making the bino heavier vis-a-vis the
Higgsinos.  Taking $M_1\ga M_2$ at the electroweak scale roughly doubles
the size of the Higgsino regions in Fig.~\ref{fig:mumd} and more than
doubles the allowed areas in Fig.~\ref{fig:area}.

We have shown that while non-universal Higgs mass parameters allow for
a Higgsino-type neutralino in models with gaugino mass unification, it
is at the price of living in a narrow strip in the $\{\mhu,\mhd\}$
parameter space.  This fact is often obscured by scatter plots which
tend to emphasize the maximum extent to which the low energy
parameters can be affected by GUT scale non-universalities, rather
than the likelihood of the resulting parameters.  In this work we have
considered only non-universalities in the Higgs masses.  Extending
scalar mass non-universality to the sfermions can have a significant
impact on the running of $\mhu$ via the D-term contribution $S$.
However, it is likely that more tuning would be required to cancel the
$S$ term contribution to (\ref{eq:rge}) with the Yukawa terms and
permit a small $|\mu|$.  Thus a gaugino-type neutralino is still
preferred in SUSY models with gaugino mass unification, even in the
absence of scalar, in particular Higgs, mass unification.

\vskip .2in
\vbox{
\noindent{ {\bf Acknowledgments} } \\
\noindent  I would like to thank Vernon Barger for helpful
discussions and comments.  This work was supported in part by DOE   
grant DE--FG02--95ER--40896 and in part by the University of Wisconsin  
Research Committee with funds granted by the Wisconsin Alumni Research  
Foundation.}


\vspace{-.1in}
\begin{thebibliography}{99}
\bibitem{tev2k} TeV2000 mSUGRA Working Group Report,\hfill\\
 {\tt http://ftp-pheno.physics.wisc.edu/sugra.ps.gz}; \hfill\\ H. Murayama and 
 M. Peskin, Ann. Rev. Nucl. Part. Sci. {\bf 46} (1996) 533.
  
\bibitem{efgos} J. Ellis, T. Falk, G. Ganis, K.A. Olive and M.
  Schmitt, Phys. Rev. {\bf D58} (1998) 095002.
  
\bibitem{ikkt}K. Inoue, A. Kakuto, H. Komatsu and S. Takeshita,
  Prog. Theor. Phys., Vol. 68, No. 3 (1982) 927.

\bibitem{mv}S.P. Martin and M.T. Vaughn, Phys. Rev. {\bf D50} (1994)
  2282.

\bibitem{pp}D. Pierce and A. Papadopoulos,
Phys. Rev. {\bf D50} (1994) 565 and Nucl. Phys. {\bf B430} (1994) 278.

\bibitem{bbo}V. Barger, M. S. Berger and P. Ohmann, Phys. Rev.  {\bf
    D49}(1994) 4908.

\bibitem{pbmz}D. Pierce, J. Bagger, K. Matchev and R. Zhang,
  Nucl.Phys. {\bf B491}(1997)3.
   
\bibitem{dn}M. Drees and M. M. Nojiri Nucl. Phys. {\bf B369} (1992)
  54 .

\bibitem{popo}N. Polonsky and A. Pomarol, Phys. Rev. Lett. {\bf 73}
  (1994) 2292; Phys. Rev. {\bf D51} (1995) 6532; Y. Kawamura,
  H. Murayama and M. Yamaguchi, Phys. Rev. {\bf D51} (1995) 1337.
  
\bibitem{mop}H. Murayama, M. Olechowski and S. Pokorski, Phys. Lett.
  {\bf B371} (1996) 57.

\bibitem{tuning} J. Ellis, K. Enqvist, D. Nanopoulos and F. Zwirner,
  Nucl. Phys. {\bf B276} (1986) 14; R. Barbieri and G.F. Giudice,
  Nucl. Phys. {\bf B306} (1988) 63.
   
 \bibitem{anderson}G. W. Anderson and D. J. Castano, Phys. Lett. {\bf
     B347} (1995) 300; G. W. Anderson and D. J.  Castano, Phys. Rev.
   {\bf D52} (1995) 1693; G. W. Anderson and D. J.  Castano, Phys.
   Rev. {\bf D53} (1996) 2403.
 
\bibitem{op1}M. Olechowski and S. Pokorski, Phys. Lett. {\bf
     B344} (1995) 201.
 
 \bibitem{tuning2}P. Chankowski, J. Ellis, and S.  Pokorski,
   Phys.  Lett. {\bf B423} (1998) 327; P. Chankowski, J.  Ellis, M.
   Olechowski and S. Pokorski, hep-ph/9808275; R. Barbieri and A.
   Strumia, Phys.  Lett. {\bf B433} (1998) 63; P. Chankowski, J.
   Ellis, K. A. Olive and S. Pokorski, hep-ph/9811284.
  
\bibitem{nonu} S. Dimopoulos and G.F.  Giudice, Phys. Lett. {\bf B357} (1995) 573.

\bibitem{ccn} Kwok Lung Chan, Utpal Chattopadhyay, and Pran
  Nath, Phys.Rev. {\bf D58} (1998) 096004.

\bibitem{wright}D. Wright, hep-ph/9801449.

\bibitem{kk} G.L. Kane and S.F. King, hep-ph/9810374.

\bibitem{cos}V. Berezinskii , A. Bottino, J. Ellis, N. Fornengo,
  G. Mignola, S. Scopel, Astropart.Phys. {\bf 5} (1996) 1;
P. Nath and R. Arnowitt, Phys.Rev. {\bf D56} (1997) 2820.


\end{thebibliography}
\end{document}